\documentclass{article}

\usepackage[english]{babel}

\usepackage[letterpaper,top=2cm,bottom=2cm,left=3cm,right=3cm,marginparwidth=1.75cm]{geometry}
\usepackage[plain,noend]{algorithm2e}
\newcommand{\detm}{\operatorname{det}}
\newtheorem{theorem}{Theorem}

\usepackage{natbib}
\usepackage{amsmath}
\usepackage{graphicx}
\usepackage{amsmath}
\usepackage{amsfonts}
\usepackage{subcaption}
\usepackage[colorlinks=true, allcolors=blue]{hyperref}
\usepackage{multirow} 
\usepackage{booktabs} 
\usepackage{subcaption}

\usepackage{setspace}
\title{Modeling Non-Uniform Hypergraphs Using Determinantal Point Processes}

\author{
  Yichao Chen\thanks{Email: \texttt{yichaoc@umich.edu}} \\
  Department of Statistics, University of Michigan, Ann Arbor, Michigan \\[1ex]  
  Jingfei Zhang\thanks{Email: \texttt{emma.zhang@emory.edu}} \\
  Goizueta Business School, Emory University, Atlanta, Georgia \\[1ex]
  Ji Zhu\thanks{Email: \texttt{jizhu@umich.edu}} \\
  Department of Statistics, University of Michigan, Ann Arbor, Michigan
}

\date{}
\begin{document}

\maketitle
\begin{abstract}
Most statistical models for networks focus on pairwise interactions between nodes. However, many real-world networks involve higher-order interactions among multiple nodes, such as co-authors collaborating on a paper. Hypergraphs provide a natural representation for these networks, with each hyperedge representing a set of nodes. The majority of existing hypergraph models assume uniform hyperedges (i.e., edges of the same size) or rely on diversity among nodes. In this work, we propose a new hypergraph model based on non-symmetric determinantal point processes. The proposed model naturally accommodates non-uniform hyperedges, has tractable probability mass functions, and accounts for both node similarity and diversity in hyperedges. For model estimation, we maximize the likelihood function under constraints using a computationally efficient projected adaptive gradient descent algorithm. We establish the consistency and asymptotic normality of the estimator. Simulation studies confirm the efficacy of the proposed model, and its utility is further demonstrated through edge predictions on several real-world datasets.

\end{abstract}

\smallskip
\noindent \textbf{Keywords:} hypergraphs; higher-order interactions; determinantal point process; nonuniform hyperedges; constrained estimation.

\newpage
\section{Introduction}

Existing network analyses have primarily focused on pairwise interactions, where each edge in the network consists of two nodes. However, higher-order interactions, which involve multiple nodes simultaneously, are ubiquitous across many real-world scenarios. For example, in academic collaborations, researchers often co-author papers in teams of three or more. In protein-protein or gene-gene interactions, multiple proteins or genes may work as a group to carry out biological functions. Existing studies have demonstrated the importance of higher-order interactions in contexts such as neural systems \citep{yu2011higher}, genetic networks \citep{ritz2014signaling}, and the spread of epidemics \citep{battiston2021physics}. Despite their importance, these high-order interactions are often simplified into pairwise interactions in existing analyses \citep{grover2016node2vec, cao2015grarep}, resulting in substantial information loss. For example, in co-author networks, a paper co-authored by three or more represents a collaborative effort among all co-authors. Reducing this high-order interaction to pairwise connections fails to capture the full information of the collaboration.
A more natural approach is to directly model higher-order interactions, allowing for richer and more accurate representations of complex systems.

High-order interactions among a set of nodes can be naturally represented using \textit{hypergraphs}, a generalization of traditional graphs where an edge, known as a \textit{hyperedge}, is a set that includes all interacting nodes. For example, in a co-author network, nodes represent authors and a hyperedge corresponds to the set of authors of a paper. Recent studies have explored hypergraphs through various approaches. One line of work transforms hypergraphs into weighted graphs using hypergraph Laplacians \citep{Ghoshdastidar_Dukkipati_2017, chien2018community}, where edge weights are based on the counts of hyperedges shared by each node pair. 
Other works focus on simplicial hypergraphs, which require that the presence of a hyperedge implies the presence of all its subsets \citep{lunagomez2017geometric,bobrowski2018topology}. Another line investigates uniform hypergraphs, where all hyperedges contain the same number of nodes \citep{ghoshdastidar2017uniform, ke2019community, Yuan_Qu_2022, lyu2023latent}. Extensions to non-uniform hypergraphs have been proposed by either augmenting hyperedges with null vertices to create uniform multi-hypergraphs \citep{zhen2023community} or by combining multiple uniform hypergraphs of different orders \citep{yuan2022testing, turnbull2023latent, nandy2024degree}.  The majority of these works focus on network community detection and represented hypergraphs using tensors. However, tensor-based representations can be computationally inefficient, as the order of the tensor is determined by the hyperedge with the largest cardinality, leading to significant scalability challenges. They also cannot accommodate multi-hyperedges. That is, two or more hyperedges connecting the same set of nodes, which can occur, for example, when one group of co-authors collaborated on several papers. 
Notably, \cite{Xianshi2022} proposed a flexible non-uniform latent space hypergraph model that accounts for multi-hyperedges. Their model naturally accommodates non-uniform hyperedges and has a tractable probability mass function. In particular, their model is defined via a class of determinantal point process, a discrete probabilistic model parameterized by a kernel matrix. The proposed model assigns a probability to every subset of nodes based on the diversity of the included nodes, and nodes with diverse latent positions are more likely to form hyperedges.
While assuming diversity can be plausible in some settings, such as ingredients in a recipe, similarity, where similar nodes may more likely to appear in an hyperedge, can be more plausible in many other settings, such as co-authors on a paper.

In our work, we propose a new hypergraph model using a more flexible class of determinantal point processes. The proposed model naturally accommodates non-uniform hyperedges, has tractable probability mass functions, and allows for node similarity or diversity in hyperedges. 
In particular, we consider a nonsymmetric kernel matrix, which overcomes the restrictive negative association property induced by symmetric kernels \citep{brunel2018learning} and represents a major improvement over \cite{Xianshi2022}.
For model estimation, we maximize the likelihood function under constraints via a computationally efficient projected adaptive gradient descent algorithm. For theoretical results, we show the consistency and the asymptotic normality for the constrained maximum likelihood estimators. The special manifold of the parameter space of our model leads to a challenging proof, especially for the asymptotic normality. %Compared with \cite{Xianshi2022}, we use more complicated tools to verify several assumptions required for achieving the asymptotic normality. 
Notably, the asymptotic distribution is intricately linked to a tangent cone with a unique structure. The nonsymmetric kernel matrix in our model results in a unique tangent cone, requiring very delicate derivations to derive asymptotic normality.

\section{Model}
\subsection{Notation}
Write $[k]=\{1,2, \ldots, k\}$. Let $\|\cdot\|_2$ denote the vector $\ell_2$-norm, $\|\cdot\|_F$ the matrix Frobenius norm and $\detm(\cdot)$ the determinant of a matrix. 
Denote the $i$th row and $j$th column of a matrix $M$ as $M_{i .}$ and $M_{. j}$, respectively. Let $\operatorname{vec}(M)$ denote the vector formed by concatenating the columns of matrix $M$, and $\operatorname{diag}(M)$ denote the vector with diagonals of matrix $M$. Let $\operatorname{vec}^{-1}(\cdot)$ be the inverse function of $\operatorname{vec}(\cdot)$ such that  $\operatorname{vec}^{-1}\left(\operatorname{vec}(M)\right) =M$. Write $H_n$ as the set of $n \times n$ diagonal matrix with diagonal elements in $\{-1,1\}$, and $O_d$ as the set of $d \times d$ orthogonal matrices. Given a $n \times n$ matrix $M$, denote $M_{e_s}$ as the $|e_s|\times |e_s|$ principal submatrix of $M$ indexed by $e_s\subset[n]$.

%\jz{[Yichao, for this section, give an introduction of NDPP directly and state DPP afterwards as its special case. To be more specific, first introduce equation (1), which holds with or without the DPP assumption. Then state the sufficient condition on $L$ for defining a proper probability distribution over all subsets. After this, you can state DPP as a special case of $L$ and the implication of having a symmetric $L$.]} \yc{[modified]}

\subsection{Determinantal Point Process}
\label{sec:DPP}
%To model hypergraphs and overcome the bottleneck of computational costs, we hope to construct a tractable distribution with accessible normalizing constants, which motivates us to adopt the results for determinantal point processes(DPPs), which is one special discrete, finite point process that can lead to normalizing constants. We first give the formal definitions about the DPP and then we will discuss the connection between the DPP and the hypergraph modeling.
%Determinantal point processes were first introduced in \cite{macchi1975coincidence}, where they were referred to as ``fermion processes" because they represent the distributions of fermion systems in thermal equilibrium. 
A point process on a set $V$ is a probability measure defined over finite subsets of $V$. In the case of discrete, finite point processes, where $V = \{1, 2, \ldots, n\}$, a point process on $V$ is a probability measure over all subsets of $V$. A point process on $V$ is called a \textit{determinantal point process} if, for a randomly drawn subset $E\subseteq V$ and a given subset $e \subseteq V$, the following probability equation holds for an $n \times n$ matrix $L$:
\[
pr_L(E = e) \propto \operatorname{det}\left(L_e\right).
\]
%where $L$  is a $n \times n$ matrix indexed by the elements of $V$ whose principal minors are nonnegative.
%where $L$  is real, symmetric positive semidefinite $n \times n$ matrix indexed by the elements of $V$. 
%Here, the submatrix $L_e = \left[L_{i j}\right]_{i, j \in e}$ denotes the restriction of $L$ to the entries indexed by the elements of $e$.  
By the property of matrix determinant, it can be shown that $\sum_{e \subseteq V} \operatorname{det}\left(L_e\right)=\operatorname{det}(L+I)$, where $I$ is the identity matrix. 
Correspondingly, the following holds 
\begin{equation}
	pr_L(E=e)=\frac{\operatorname{det}\left(L_e\right)}{\operatorname{det}(L+I)}.
	\label{DPP_def}
\end{equation}
%From Equation \eqref{DPP_def}, for the hyperedge $e_s, \ s = 1,\ldots , m$, we have
%$$
%	pr(e_s)=\frac{\operatorname{det}\left(L_{e_s}\right)}{\operatorname{det}(L+I)}
%$$
The matrix $L$ is referred to as a \textit{kernel matrix}. Not all matrices $L$ are admissible to define a determinantal point process. A matrix $L$ is admissible if and only if all principal minors of $L$ are nonnegative \citep{brunel2018learning}; such matrices are called $P_0$-matrices.

The majority of existing literature on determinantal point processes assume $L$ to be symmetric and positive semidefinite \citep{kulesza2012determinantal, lavancier2015determinantal,kang2013fast}, which is a special case of $P_0$-matrices. Interestingly, assuming symmetry on $L$ induces a strong property of negative dependence called \textit{negative association} \citep{brunel2018learning}. To explain this property, we first introduce the marginal kernel matrix $K$, where $K = I  - (L + I)^{-1}$. For a given subset $e \subseteq V$ and a randomly drawn subset $E\subseteq V$, it holds under \eqref{DPP_def} that
$$
pr( e \subset E) = \operatorname{det}(K_e).
$$
It is seen that $K$ models the marginal probabilities of subsets. If $e = \{i\}$, then $pr(i \in E) = K_{ii}$. 
When $L$ is symmetric, $K$ is also symmetric. Under symmetric kernels one can show that $\text{Cov}\{1(i\in E),1(j\in E)\}=-K_{ij}^2$ for any $i\neq j$. More generally, for any two disjoint sets $e_1,e_2\in V$, it holds that  $\text{Cov}\{1(e_1\subseteq E),1(e_2\subseteq E)\}=\operatorname{det}(K_{e_1\cup e_2})-\operatorname{det}(K_{e_1})\operatorname{det}(K_{e_2})\le 0$ \citep{borcea2009negative}. 
%\jz{[Yichao, I saw this in \citep{brunel2018learning}. Could you verify that this is true?]} \yc{[Thank you for adding this! Yes, I think it is true. The inequaility comes from the negative association.]}
This negative association from symmetric kernels forces repulsive interactions between items in $V$.

\subsection{A New Non-uniform Hypergraph Model}
Let $V= \left[n\right]$ denote the set of $n$ nodes, 
and $D = \left\{e_1, e_2, \ldots, e_{m}\right\}$ the set of $m$ observed hyperedges. 
Each hyperedge can be represented as a subset of $V$, that is, $e_s \subset \left[n\right]$ for $s\in[m]$. 
We assume that $e_s$ for $s\in[m]$ follows
$$
e_s \stackrel{\text { i.i.d. }}{\sim} pr_{L}, 
$$
where $L$ is a kernel matrix and $pr_{L}$ is as defined as \eqref{DPP_def}. Modeling hyperedges using determinantal point processes has several benefits. First, it naturally accommodates non-uniform hyperedges and multi-hyperedges, which greatly extends model flexibility. Second, as the normalizing constant can be easily derived, $pr_{L}$ defines a tractable distribution over all possible hyperedges, facilitating estimation, inference, and sampling. Third, the kernel matrices $L$ and $K$ enhance model interpretability as discussed in Section \ref{sec:DPP}.

%While assuming diversity can be plausible in some settings, such as ingredients in a recipe, similarity, where similar nodes may more likely to appear in an hyperedge, can be more plausible in many other settings, such as co-authors on a paper. 

To introduce more flexibility when forming hyperedges, we consider a general form of kernel matrices that are not necessarily symmetric. Since the kernel matrix $L$ needs to be a $P_0$-matirx, we choose $L$ such that $L + L^T $ is a positive semidefinite matrix, which is a sufficient condition for $L$ to be a $P_0$-matrix.
We propose the following nonsymmetric determinantal point process hypergraph model:
\begin{align}
\label{L_equation}
&pr_L(E=e)=\frac{\operatorname{det}\left(L_e\right)}{\operatorname{det}(L+I)}, \\\nonumber
&L = B V V^T B + \gamma \cdot BV C V^T B + B^2,
\end{align}
where $B$ is a $n \times n$ diagonal matrix whose diagonal elements $\beta_i$'s are non-negative and characterize node popularity. The matrix $V$ is a $n \times d$ with the constraint $\|v_i\|_2 = 1$, where $v_i$ denotes the $i$th row of $V$. The vector $v_i$ can be viewed as the latent space position for the $i$th node (See Equation \ref{eq:two node cross product}). The matrix $C$ is a skew matrix satisfying $C^T = -C$, with the constraint that $\|C\|_F = 1$. This special form of $C$ is chosen to guarantee that $L + L^T$ is a positive semidefinite matrix. %so that we have a closed-form probability for each possible hyperedge in the hypergraph. 
The scalar $\gamma\ge 0$ is a scaling parameter for the skew matrix $C$.
The extra term $B^2$ is introduced to ensure $L$ is full-rank. 
If $L$ is rank-deficient, then $pr_L(E=e)=0$ for $|e|>$ rank($L$), as $\detm (L_e)=0$. We write the hypergraph model in \eqref{L_equation} as $H(L)$ or $H(B, V, C, \gamma)$.

%Otherwise, suppose $L$ is a matrix of rank-$d$, when the size of hyperedge $e$ is greater than $d$, i.e.,$|e| > d$, the corresponding determinant  $\detm (L_e)$ will be zero and the probability of generating hyperedges with the sizes larger than $d$ will be zero. We can demonstrate the flexibility of our hypergraph model using NDPPs by checking the probability of $e_l$ when $|e_l| = 2$, i.e. the hyperedge only consists of two nodes :
To help understanding \eqref{L_equation}, we consider a two-element set $\{i,j\}$, which gives:
\begin{equation}
pr_L(E = \{i,j\}) \propto 4\beta_{i}^2 \beta_{j}^2 - \beta^2_{i}\beta^2_{j} \cos^2(v_i, v_j) + \gamma^2 \beta_i^2 \beta_j^2 (v_i^T C v_j)^2.
\label{eq:two node}
\end{equation}
When $\beta_i$'s are fixed and $\gamma = 0$, the probability $pr_L(E = \{i,j\})$ is driven by $-\cos^2(v_i, v_j)$. Correspondingly, this model encourages diversity, that is, nodes with unaligned latent positions are more likely to form hyperedges. 
%This is similar to the model considered in \cite{Xianshi2022}. 
%As $v_i$ and $v_j$ move further apart from each other, the probability increases.
When $\gamma\neq0$, the term $(v_i^T C v_j)^2$ may encourage nodes with similar latent positions to form hyperedges. For example, set $d = 3$ and consider $C$ in the form of
$$
C=\left[\begin{matrix}
0 & \lambda_3 & -\lambda_2 \\
-\lambda_3 & 0 & \lambda_1 \\
\lambda_2 & -\lambda_1 & 0
\end{matrix}\right],
$$
where $\lambda_1, \lambda_2, \lambda_3 \in R$. The matrix $C$ is determined by the vector $\lambda = (\lambda_1, \lambda_2, \lambda_3)$. By properties of skew-symmetric matrices, we have $v_i^T C = (\lambda  \times v_i)^T$ where $\lambda \times v_i$ represents the cross product between $\lambda$ and $v_i$. That is, we have
\begin{equation}
P(E = \{i,j\}) \propto 4\beta_{i}^2 \beta_{j}^2 - \beta^2_{i}\beta^2_{j} \cos^2(v_i, v_j) + \gamma^2 \beta_i^2 \beta_j^2 \cos^2(\lambda \times v_i, v_j).
\label{eq:two node cross product}
\end{equation}

\begin{figure}[!t]
\centering
\begin{subfigure}{.48\textwidth}
  \centering
  \includegraphics[width=\linewidth]{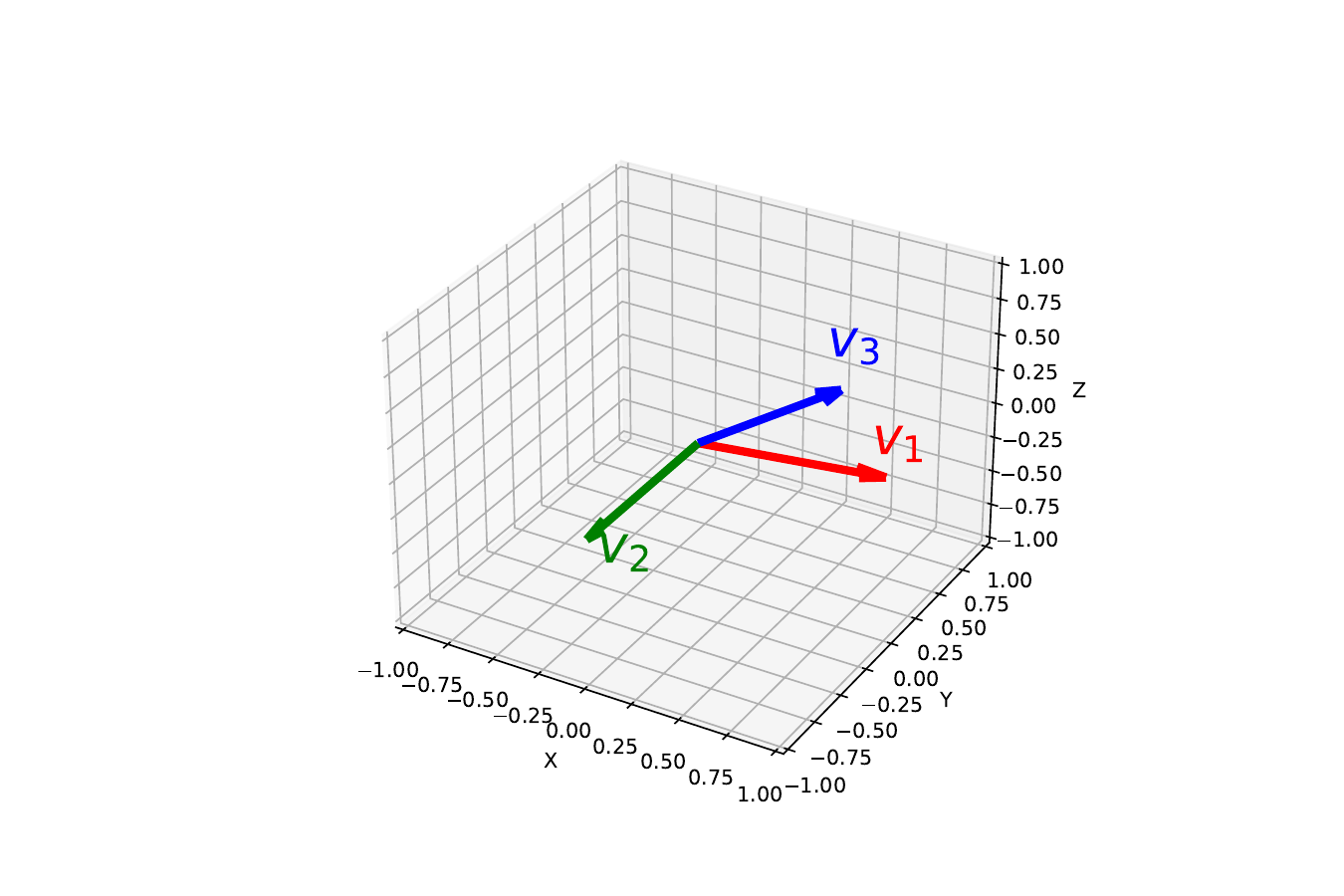}  
  \caption{ Latent positions for $v_1, v_2, v_3$}
  \label{latent position}
\end{subfigure}%
\hfill
\begin{subfigure}{.48\textwidth}
  \centering
  \includegraphics[width=\linewidth]{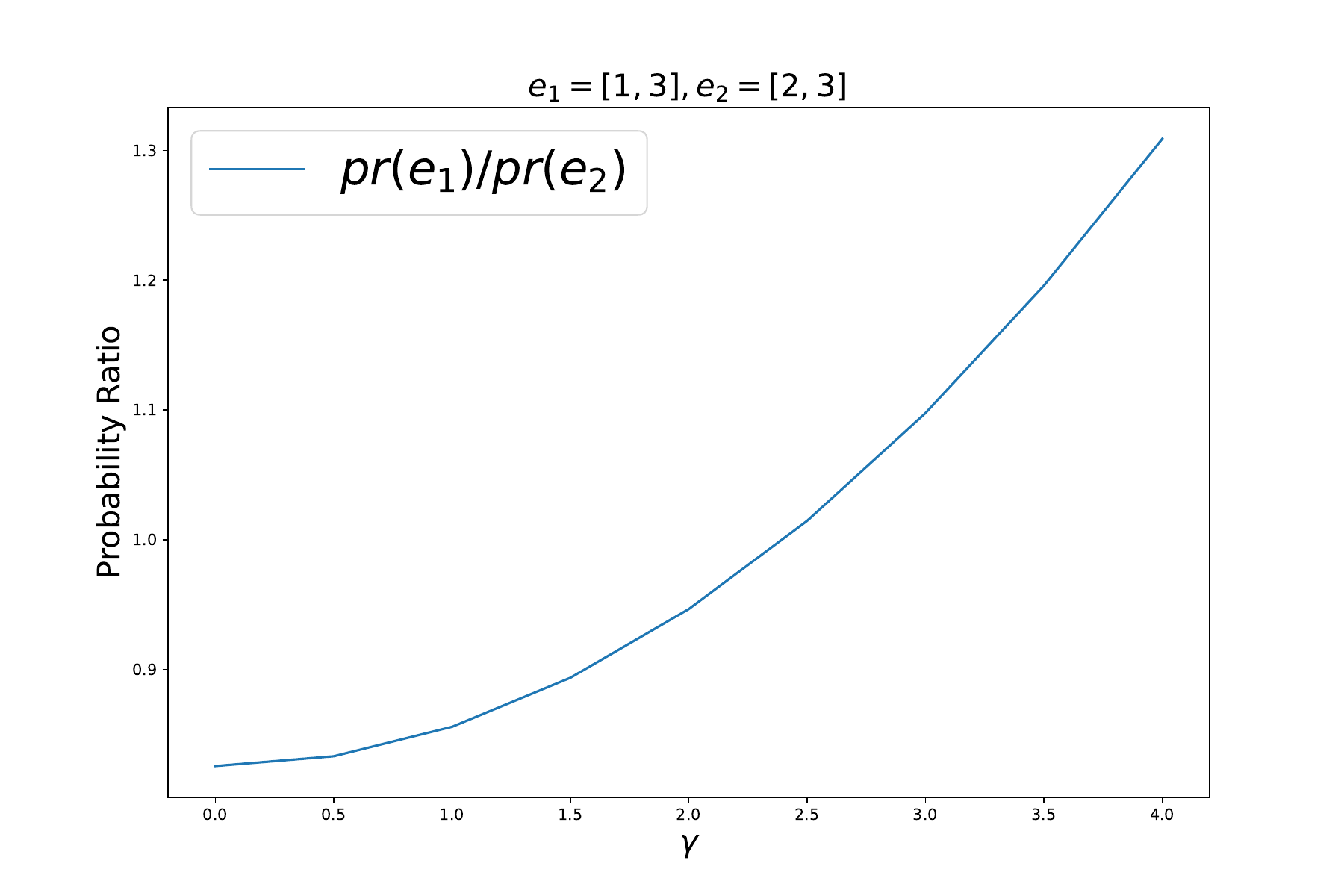}
  \caption{The changes of probability ratio as $\gamma$ changes}
  %\vspace{-50pt}
  \label{probability ratio}
\end{subfigure}
\caption{An illustration of the non-symmetric determinantal point process hypergraph model.}
\label{fig:NDPP_change_gamma}
\end{figure}
The direction of vector $\lambda \times v_i$ is perpendicular to both $\lambda$ and $v_i$, and the magnitude of vector $\lambda \times v_i$ is the area of the parallelogram spanned by $\lambda$ and $v_i$. The cross product $\lambda \times v_i$ leads to the position change for a node latent vector $v_i$ and two diverse nodes may form hyperedges with smaller probabilities. In addition, as we increase the value of $\gamma$, the term $\gamma^2 \beta_i^2 \beta_j^2 \cos^2(\lambda \times v_i, v_j)$ contributes more to the probability. In summary, $v_i$'s, $\gamma$ and $\lambda$ together modulate the probabilities of forming hyperedges. 

%We give a demonstration example with randomly chosen vectors $v_1, v_2, v_3$ and $\lambda$. 
In Figure \ref{fig:NDPP_change_gamma}, we present an illustrative example with latent positions vectors $v_1, v_2, v_3$ and the probabilities of $e_1 = \{1,3\}$ and $e_2 = \{2,3\}$ with varying values of $\gamma$. Since the latent positions $v_1$ and $v_3$ are more similar than $v_2$ and $v_3$, we have $pr(e_1)/ pr(e_2) < 1$ when $\gamma = 0$, corresponding to a symmetric kernel and encourages diversity in the generated hyperedges.  
As $\gamma$ increases, the ratio of $pr(e_1)/ pr(e_2)$ increases monotonically. When $pr(e_1)/pr(e_2) > 1$, similar positions $v_1$ and $v_3$ are more likely to form hyperedges compared to diverse positions $v_1$ and $v_2$. This example demonstrates the extra flexibility provided by the non-symmetric kernel matrix $L$ in our model. %Either the similarity or the diversity among nodes can be allowed to accommodate the nature of data. 

\subsection{Model Properties and Sampling}
Based on model \eqref{L_equation}, we can derive the following useful properties on the marginal and conditional distributions. 

\medskip
\noindent \textbf{Marginal Distribution.}
Recall $K = I - (L+I)^{-1}$. Under \eqref{L_equation}, the marginal probability for node $i$ to be included in a random hyperedge is
$$
pr(i \in E)=K_{i i} .
$$

\noindent \textbf{Conditional Distribution.} For any $e_1 \subset e_2 \subset\left[n\right]$ and under \eqref{L_equation}, we have
\begin{equation}
pr\left(E =e_2 \mid e_1 \subset E\right)=\frac{\operatorname{det}\left(L_{e_2}\right)}{\operatorname{det}\left(L+I-I_{e_1}\right)}
\label{eq: condition distribution}
\end{equation}
where $I_{e_1}$ is a $n \times n$ diagnonal matrix with $\{I_{e_1}\}_{ii} = 1 $ if $i \in e_1$, and 
$\{I_{e_1}\}_{ii} = 0$ otherwise.

\medskip
\noindent \textbf{Invariant to node ordering.} Under \eqref{L_equation}, the probability of forming a hyperedge $e$ is invariant to the ordering of nodes in the hyperedge $e$ and the node set $V$. This follows from the determinant's properties. Any reordering of nodes can be achieved through a sequence of pairwise swaps. Swapping two nodes requires swapping two rows and two columns in the corresponding kernel matrix. According to Section 5.3 of \cite{hoffmann1971linear}, each such swap affects only the sign of the determinant without altering its absolute value. Consequently, swapping two nodes does not change the values of $\operatorname{det}(L_e)$ and $\operatorname{det}(L+I)$, ensuring that the probability of forming hyperedges remains invariant to node ordering.

\iffalse
\noindent \textbf{Disbution of the cardinality of $\boldsymbol{E}$.} 
The cardinality of the random hyperedge is distributed as the sum of $n$ independent Bernoulli trials with the success probabilities respectively being $p_k$ with  $p_k = \lambda_k(L) /\left(1+\lambda_k(L)\right)\left(k=1, \cdots, n\right)$, where $\lambda_k(L)$ is the $k$ th eigenvalue of $L$. In particular, we have
$$
\mathbb{E}(|E|)=\sum_{k=1}^{n} \frac{\lambda_k(L)}{1+\lambda_k(L)} \text { and } \operatorname{Var}(|E|)=\sum_{k=1}^{n} \frac{\lambda_k(L)}{\left(1+\lambda_k(L)\right)^2}
$$
\fi
%Since $L+L^T$ is a positve semidefinite matrix, the proof of the properties remains the same as DPP given all the eigenvalues of $L$ are non-negative and Equation \eqref{DPP_def} still hold true.
Given model parameters $B, V, C, \gamma$, hyperedges can be directly sampled from \eqref{L_equation} using the following algorithm 

\begin{algorithm}
\SetAlgoLined
Set  $Z = (B^{2}+I)^{-1} B V$, $W = ((I_d + \gamma C)^{-1} + V^T B (B^2 + I_n)^{-1} B V)^{-1}$, $Y = \emptyset$\; 
\For{$i = 1$ to $n$}{
$p_i \gets {z_i}^T W z_i$\;
$u \gets \text{uniform}(0,1)$\;
if $u \leq p_i$, then $Y \gets Y \cup \{i\}$; else $p_i \gets p_i - 1$ \;
$W \gets W - \frac{W z_i z_i^T W}{p_i}$\;
\Return Y
}
\caption{Hyperedge Sampling from \eqref{L_equation}}
\label{NDPP sampling}
\end{algorithm}

Algorithm \ref{NDPP sampling} is adapted from Theorem 2 in \cite{poulson2020high}, which provides a factorization-based  sampling method for generic determinantal point processes. To sample a hyperedge, we initialize it as an empty set and iterate through each node to decide if it should be included based on decisions for previous nodes. The inclusion/exclusion decisions need to be made with the appropriate conditional probabilities, which can be calculated based on the properties of model \eqref{L_equation}. %NDPP 

\subsection{Identifiability}
For model \eqref{L_equation}, it can be shown that the matrix $V$ is identifiable up to an orthogonal transformation and sign flips for rows, and the matrix $C$ is identifiable up to orthogonal transformation on both sides. % for both left and right sides, i.e., with any orthogonal matrix $O$, the two hypergraphs $H(V, B, C, \gamma)$ and $H(SVO^T, O C O^T, B, \gamma )$  are equivalent. 
The formal results are presented in Theorem \ref{thm:identifiability}, and its proof is collected in the supplement. 

\begin{theorem}
\label{thm:identifiability}
Suppose  $S \in H_n$, $O \in O_{d}$ and matrix $V$ is full rank. For two different kernel matrices $L_1 = B_1 V_1{V_1}^T B_1  + \gamma_1 B_1 V_1 C_1 {V_1}^T B_1 + {B_1}^2 $ and $L_2 = B_2 V_2{V_2}^T B_2  + \gamma_2 B_2 V_2 C_2 {V_2}^T B_2 + {B_2}^2$, $L_1$ and $L_2$ give equivalent hypergraph models if $L_1 = S L_2 S$. 
%\jz{[Yichao, is this ``if and only if"?]}\yc{[I cannot prove the other side, only ``if'' here]}
Furthermore, we have $L_1 = S L_2 S$ if and only if $V_1 = S V_2 O^T$, $C_1 = O C_2 O^T$, $B_1 = B_2$ and $\gamma_1 = \gamma_2$.
\end{theorem}

\subsection{Parameter Estimation}

Given an observed set of hyperedges $D = \{e_1, \ldots e_{m}\}$, we aim to estimate parameters $B, V, C, \gamma$. Under \eqref{L_equation}, the likelihood function can be written as 
\begin{align*}
&\phi(B, V, C, \gamma)= \\
&\frac{1}{m} \sum_{s=1}^{m} \log \operatorname{det}\left(B_{e_s} V_{e_s}(I+\gamma C) V_{e_s}^T B_{e_s}+B^{2}_{e_s}\right)-\log \operatorname{det}\left(B V(I+\gamma C) V^T B+B^2+I\right).
\end{align*}
We consider the following constrained optimization problem:
\begin{align*}
&\underset{B, V, C, \gamma}{\operatorname{max}} \quad \phi(B, V, C, \gamma) \\
&\text{s.t} \quad  \operatorname{diag}(B) > 0, \,\, \operatorname{diag}(V V^T) = 1, \,\, \|C\|_F = 1.
\end{align*}

This objective function is nonconvex, and we propose to solve it via an efficient projected adaptive gradient descent algorithm. 
The detailed algorithm is provided in the supplement.  
The adaptive gradient descent algorithm was first proposed in \citep{kingma2014adam}, and is usually referred to as the Adam algorithm. 
The learning rate is adaptively chosen and gradually reduced until the loss has stopped improving. The Adam algorithm combines desirable features of Momentum 
 \citep{qian1999momentum} and RMSprop\citep{tieleman2012lecture}. %\jz{[Yichao, add references to these two algorithms.]} \yc{[added]} 
 Compared with a standard project gradient descent, it can provide adaptive learning rates, faster convergence, and robustness to the choice of hyperparameters. It has been proven to work well in many non-convex optimization problems.
%\jz{[Yichao, can you explain the Adam algorithm using a few sentences? We also want to discuss why we used Adam as opposed to the standard project gradient descent.]}\yc{[added]}
When implementing this algorithm, multiple random initial values are chosen and we select the results with the maximum objective value. We use the cross-validation approach to select $d$ in real-world data analysis. 
%If the loss improvement is less that 1e-5 for a total of 300 iterations, we determine the training process as converged and stop the iteration. The algorithm is implemented in PyTorch. 
%The following Algorithm \ref{Adam alg} provides the details for our optimization algorithms. 
%\jz{[Did we discuss at all how $d$ is selected?]}\yc{[in real data sections, use validation approach to choose d, elsewhere I just set d = 3.]}

\section{Theoretical results}
This section establishes the consistency and asymptotic normality of the MLE estimates. We first introduce some notation.
%In this section, we obtain the consistency and the asymptotic normality of our MLE estimates for model parameters. We observe $m$ different hyperedges $\{e_1, \ldots, e_{m} \}$ from the hypergraph, and our goal is to estimate the pamameters $(B, V, C, \gamma) $ as proposed in our hypergraph model. The kernel matrix $L$ associated with the NDPP is constructed based on $(B, V, C, \gamma) $. We first introduce some definitions before stating the main theorems.  Define the function $\operatorname{vec}$ as: $\operatorname{vec}: R^{n \times n} \rightarrow R^{n_{v}^2}$. For a matrix $M \in R^{n \times n} $, $\operatorname{vec}(M)$ is is the long column vector formed by concatenating the columns of the matrix $M$. %Set the upper bound for the scalar parameter $\gamma$ as $\gamma \leq T$ where $T$ is a finite scalar. 
Define the parameter space for $\omega=(V,B,C,\gamma)$ as  
$$
\begin{aligned}
\Omega  = & \{ V \in R^{n \times d}, B\in R^{n \times n}, C\in R^{d \times d}, \gamma \in R \mid \|v_i \|_2 = 1 \text{ for all }i,  \\
& \|C\|_F = 1,\,\, B \text{ is diagonal with } \operatorname{diag}(B) > 0, \,\,0 \leq \gamma \leq T \},
\end{aligned}
$$
where $T$ is a finite scalar. Define function $\Upsilon: \Omega \rightarrow R^{n \times n}$ as $\Upsilon(\omega)=B V V^T B + \gamma B V  C V^T B + B^2$. Correspondingly, the parameter space for $L$ is $\Psi =  \{ \Upsilon(\omega) \mid \omega \in  \Omega \}$. Let $\bar{\Psi}$ be the closure of $\Psi$. 
Denote the true model parameters as $\omega^* = (V^*, B^*, C^*, \gamma^*)$ and the true kernel matrix as $L^*$, where $L^* = \Upsilon(\omega^*)$. The Jacobian matrix of $\Upsilon(\omega)$ at $\omega^*$ is denoted as $\nabla \Upsilon\left(\omega^*\right)$.

%Recall that $pr_{L}$ represents the probability distribution with the non-symmetric determinantal point process under the true kernel matrix $L$ over the node set $V$. For an observed hyperedge set $\{e_1,\ldots, e_{m}\}$, if we assume $e_s \stackrel{\text { i.i.d. }}{\sim} pr_{L}$ for each $s$, 
We rewrite the log-likelihood as a function of $L$ such that
$$\phi(L) = \sum_{e \subseteq[n]} \hat{p}_{e}\log \operatorname{det}\left(L_e\right)-\log \operatorname{det}(L+I),
$$ where $\hat{p}_{e}= \frac{1}{m} \sum_{s=1}^{m} I(e_s = e)$. 
Define ${p}_{e}(L)=\frac{\det({L}_{e})}{\det(L+I)}$, and we have
\begin{equation}\label{phi_def}
\begin{aligned}
\Phi(L) =E(\phi(L))&=\sum_{e \subseteq[n]} p_{e}(L^*) \log \operatorname{det}\left(L_e\right)-\log \operatorname{det}(L+I) \\
& = \sum_{e \subseteq[n]} p_{e}(L^*) \log \frac{\det (L_e)}{\det (L + I )},
\end{aligned}
\end{equation}
where the second equality holds true since $\sum_{e \subseteq[n]} p_{e}(L^*) = 1$. The theorem below shows consistency of the MLE estimates.
%Recall that $p_e(L^*) = P(E = e) = \frac{\det({L}^{*}_{e})}{\det(L+I^*)}$. Here the second equation hold true since $\sum_{e \subseteq[n]} p_{e}(L^*) = 1$. We simply write $\Phi_{L^*} = \Phi$ in the following contents.

\begin{theorem}
\label{thm:consistency}
Given an observed hypergraph on $n$ nodes with $m$ hyperedges independently generated from \eqref{L_equation}, we consider the MLE estimates $\hat{B}, \hat{V}, \hat{C}, \hat{\gamma}$ for the unknown true parameters $B^*, V^*, C^*, \gamma^*$ with $(B^*, V^*, C^*, \gamma^*) \in \Omega$. 
Let $\delta =\min _{e \subset\left[n\right]} p_{e}(L^*)$. 
Assuming
%the global optimums are obtained, and the true parameter 
$V^*$ is a full-rank matrix and $2^{n+1} e^{-\delta^2 m / 2} \rightarrow 0 $, 
%The matrix $\hat{L}$ and $L^*$ are obtained by Equation \eqref{L_equation}. 
we have, as $m \rightarrow \infty$,
\begin{align*}
&  \underset{\left\{S \in H_{n}, O \in O_{d} \right\}}{ \min } \left\|S\hat{V}O^T-V^*\right\|_F \stackrel{p}{\longrightarrow} 0,  \ \underset{\left\{O \in O_{d} \right\}}{ \min } 
\|O \hat{C} O^T - C^*\|_F \stackrel{p}{\longrightarrow} 0,  \\
& \|\hat{B} - B^*\|_F\stackrel{p}{\longrightarrow} 0,  \
\|\hat{\gamma} - \gamma^*\|_F\stackrel{p}{\longrightarrow} 0, \ \underset{\{S \in H_{n}\}}{ \min }\left\|S\hat{L}S-L^*\right\|_F \stackrel{p}{\longrightarrow} 0.
\end{align*}
\end{theorem}

The detailed proof is collected in the supplement. The proof utilizes Theorem 5.14 in \cite{van2000asymptotic} and follows a similar strategy as in \cite{Xianshi2022}. 
%\jz{[Yichao, can you give a summary of the results in Theorem 5.14 in \cite{van2000asymptotic}, such as what the result is for.]} \yc{[added]}
Theorem 5.14 in \cite{van2000asymptotic} shows that under some regularity conditions and for every compact set in the parameter space, the MLE estimator belonging to that compact set converges to the true parameter when the number of samples goes to infinity. It is a fundamental result for establishing the consistency of MLE estimators. One key step in the proof is to construct a compact set based on our parameter space that almost surely contains the estimators. Unlike \cite{Xianshi2022}, which assumes the number of nodes $n$ is fixed, our result in Theorem \ref{thm:consistency} allows $n$ to diverge as long as $2^{n+1} e^{-\delta^2 m / 2} \rightarrow 0 $. By the definition of $\delta$, this can be seen as a sparsity condition.

%\begin{remark}
%In the above theorem, we view the number of nodes $n$ in this hypergraph as finite. The result can be extended to the situation where both $n$ and $m$ are infinite. Let $\delta =\min _{e \subset\left[n\right]} p_{e}(L^*)$, if $2^{n+1} e^{-\delta^2 m / 2} \rightarrow 0 $ holds true when $m \rightarrow \infty$ and $n \rightarrow \infty$, the consistency for MLE estimates can be guaranteed. This result is obtained based on the proof for Theorem \ref{thm:consistency}. The details of proofs are shown in the appendix.
%\end{remark}

Before we establish the asymptotic normality of the MLE estimates, some more definitions are in order.

\medskip
\noindent \textbf{Definition 1}: (Matrix Irreducibility). A given matrix $M_{n \times n}$ is reducible if there exists a partition $\left\{J_1, J_2, \cdots, J_K\right\}$ of $[n]$ with $K>1$, such that $M_{i i^{\prime}}=0$ whenever $i \in J_k, i^{\prime} \in J_{k^{\prime}}$ and $k \neq k^{\prime}$. Otherwise, the matrix $M$ is said to be irreducible.
%\citep{brunel2017rates}

\medskip
\noindent \textbf{Definition 2}: (Bouligand tangent cone). The Bouligand tangent cone of a set $C\in\mathbb{R}^d$ at the point $x \in C$ is the set defined as 
%$$T_C(x)=\underset{\tau \downarrow 0}{\lim \sup } \frac{C-x}{\tau}.$$ 
$$
T_C(x)=\left\{v\in\mathbb{R}^d\mid \exists\{x_n\}\in C, \{\tau_n\}\in\mathbb{R}^+, x_n\rightarrow x, \tau_n\rightarrow 0,\frac{x_n-x}{\tau_n}\rightarrow v\right\}.
$$
A vector $v$ lies in the Bouligand tangent cone of $C$ if and only if there exist a sequence $\tau_n\rightarrow 0$ and a sequence $x_n\rightarrow x$ in $C$ such that $\left(x_n-x\right) / \tau_n \rightarrow v$. See \cite{geyer1994asymptotics} for more details.

With a slight abuse of notation, we define $\operatorname{vec}(\bar{\Psi})=\{\operatorname{vec}(M)\mid M\in \bar{\Psi}\}$, and use $T_{\operatorname{vec}(\bar{\Psi})}\left(\operatorname{vec}\left(L^*\right)\right)$ to denote the Bouligand tangent cone of $\operatorname{vec}(\bar{\Psi})$ at $\operatorname{vec}\left(L^*\right)$.

\iffalse
L space definition

Denote the feasible space for 
$L$ as $\Psi$ as follows:
\begin{align*}
\Psi= \Bigg\{ & B V V^T B + \gamma \cdot BV C V^T B + B^2 \mid  \|v_i \|_2 = 1 \text{ for all }i,
\|C\|_F = 1, \\
 & B \text{ is diagonal with } \operatorname{diag}(B) > 0, 
                     0 \leq \gamma  \leq T \Bigg\}
\end{align*}

Denote $\bar{\Psi}$ as the closure of $\Psi$. 
\fi

\begin{theorem}
\label{thm:normality}
Given an observed hypergraph on $n$ nodes with $m$ hyperedges independently generated from \eqref{L_equation} with the true kernel matrix $L^{*}$. Consider the maximum likelihood estimate $\hat{L}$ of $L^{*}$, and let $$\tilde{L}=\underset{M \in\left\{S \hat{L} S \mid S \in H_{n}\right\}}{\arg \min }\left\|M-L^*\right\|_F.$$ If $V$ is full rank and irreducible, then as $m \rightarrow \infty$ while $n$ remains fixed, we have 
$$
\sqrt{m} \cdot \operatorname{vec}\left(\tilde{L}-L^*\right) \stackrel{\text { dist }}{\longrightarrow} N\left(0, Q\left(Q^T \nabla^2 F\left(\xi^*\right) Q\right)^{-1} Q^T\right) \text { as } m \rightarrow \infty,
$$
where $\xi^*=\operatorname{vec}\left(L^*\right)$,
$
F(\xi)=-\Phi \left( \operatorname{vec}^{-1}(\xi)\right)
$,
and $Q$ is an orthogonal matrix (i.e. $Q^{T} Q=I$ ) with columns forming a basis for $T_{\operatorname{vec}(\bar{\Psi})}\left(\operatorname{vec}\left(L^{*}\right)\right)$. Here
$$
\begin{aligned}
T_{\operatorname{vec}  (\bar{\Psi})}\left(\operatorname{vec}\left(L^{*}\right)\right)  &= \operatorname{vec}\left(T_{\bar{\Psi}}(L^*)\right) = \operatorname{vec}\left( 
\nabla\Upsilon\left(\omega^*\right) T_{\Omega}\left(\omega^*\right)\right)\\
 &=  \{\operatorname{vec}\left(\nabla_{V} \Upsilon\left(\omega^*\right)X +\nabla_{B}\Upsilon\left(\omega^*\right) Y + \nabla_{C}\Upsilon\left(\omega^*\right) Z + \nabla_{\gamma}\Upsilon\left(\omega^*\right)c\right) \mid \\ 
 & \quad \quad  X \in R^{n \times d}, Y = \operatorname{diag}(M), 
M \in R^{n \times n}, Z \in R^{d \times d}, c \in R \}.
\end{aligned}
$$
It can be shown that $T_{\bar{\Psi}}\left(L^*\right)$ is a linear subspace of $R^{n \times n}$ and $T_{\operatorname{vec}(\bar{\Psi})}\left(\operatorname{vec}(L^{*})\right)$ is a linear subspace of $R^{n^2}$.
\end{theorem}
Compared with Theorem \ref{thm:consistency}, this theorem further requires that the matrix $V$ is irreducible. The proof utilizes Theorem 4.4 in \cite{geyer1994asymptotics}, which provides the asymptotic distribution for a constrained M-estimator.
%\jz{[Yichao, can you give a summary of the results in Theorem 4.4 in \cite{geyer1994asymptotics}, such as what the result is for.]}\yc{[added]}
The theoretical analysis is nontrivial, as it involves the manifold structure of our model parameters. %Compared with the proof in \cite{Xianshi2022}, it is more challenging to verify the assumptions required in Theorem 4.4 of \citet{geyer1994asymptotics}.
Compared with \cite{Xianshi2022}, we use more complicated tools to verify assumptions required for achieving the asymptotic normality. In particular, our asymptotic distribution involves a tangent cone with a special structure, and the non-symmetric kernel matrix in our model leads to a very different tangent cone that needs delicate derivations to arrive at its explicit form. 

\section{Simulations}
In this section, we conduct simulation studies to examine the finite sample performance of our proposed method. As most existing work on hypergraphs, such as \cite{Yuan_Qu_2022} and \cite{lyu2023latent}, focus on uniform hyperedges, they are not directly comparable with our method. Correspondingly, we only include the model proposed by \cite{Xianshi2022} in our comparison. We will refer to \cite{Xianshi2022} as DPP and our proposed model as NDPP. 

%To model the hypergraph which allows multi-hyperedges and non-uniform hyperedges, the only work that we notice so far is the work of \cite{Xianshi2022}. Although for traditional network which focus on pairwise interactions, many latent space models focus on generating edges based on node similarities, \cite{Yuan_Qu_2022} is the only existing work which provides latent space position for each node and encourages similarity among nodes in the hypergraph. However, they propose a very different framework which considers up to 3-order hyperedges without the allowrance of more flexible hyperedge orders. Thus, it is unrealistic to directly compare the performance of our model with others. Considering all these factors, 

\begin{figure}[!t]
    \centering
    \includegraphics[width=0.9\linewidth]{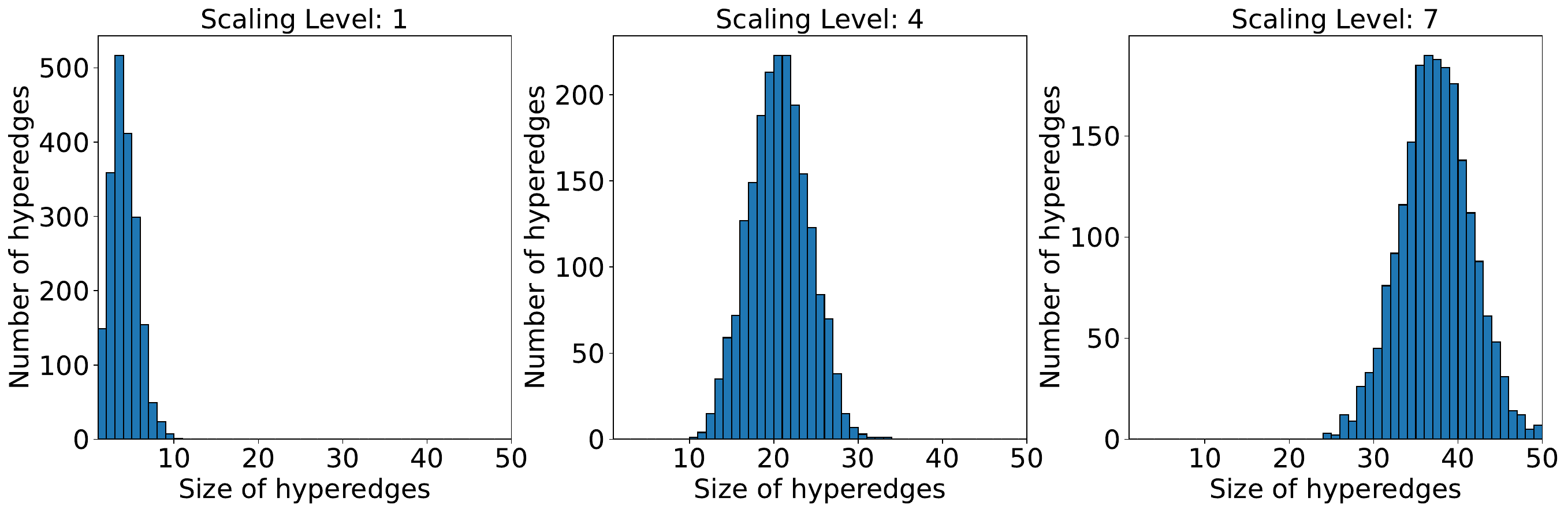}
    \caption{Histograms of hyperedge sizes under different scaling levels $s$ when the number of nodes is $n=100$ and the number of yperedges is $m = 2000$.} %\jz{[What is $m$?]\jz{[Is ``scaling level: 1" the same as $s=1$?]}}\yc{[modified]}}
    \label{fig: sparsity_demonstration}
\end{figure}

We simulate data from model \eqref{L_equation} with parameters $B, V, C,\gamma$. For $v_i$'s, elements are independently generated from Uniform$[0,1]$, and then normalized to satisfy $\|v_{i } \| = 1 $. %\jz{[what is the range for the uniform distribution?]} \yc{[modified]} 
We also consider a von Mise-Fisher (vMF) distribution, which is the spherical analogue of the normal distribution on the unit sphere. Its probability density function is given by $f(v)=C(\kappa) \exp \left(\kappa \mu^{\top} v\right)$, where $C(\kappa)$ is the normalizing constant. When generating $v_i$'s from the vMF distribution, we set $\kappa = 10$ and $\mu=(1,0,0),(0,1,0)$ and $(0,0,1)$, so that the nodes belong to three clusters. 
For $C$, we first generate a matrix $G$ with entries independently generated from Uniform[0,1], set $C = G - G^{T}$ and then rescale $C$ to ensure $\|C\|_F = 1$. The  parameter $\gamma$ is set as 0.15. 
For $B$, we generate $\eta_i \sim_{i . i . d .} \operatorname{Beta}(1,4)$ and set $\beta_i = 15s\cdot(0.2\eta_i + 0.05)$. This setting encourages only a small proportion of the nodes to have large popularity parameters. The parameter $s$ is a scaling parameter that controls the sizes of hyperedges; see Figure \ref{fig: sparsity_demonstration}. We let $s$ vary from 1 to 7. The sampling algorithm is presented in Algorithm \ref{NDPP sampling}. 
%With this choice of $\beta_i$, we make sure that a large proportion of nodes have relatively small popularity parameter values, while a small proportion of nodes will have large popularity parameter values and they are more likely to appear in hyperedges. We choose the scaling parameter $s$ from 1 to 7. It controls the sizes of hyperedges in the generated hypergraphs. The histogram for the sizes of hyperedges  with different scaling parameters is shown in Figure \ref{fig: sparsity_demonstration}. Given $V, C, \beta, \gamma$, we obtain $L$ from Equation \eqref{L_equation}. 

\begin{figure}
\hspace{2em}
 \includegraphics[width=0.9\textwidth]{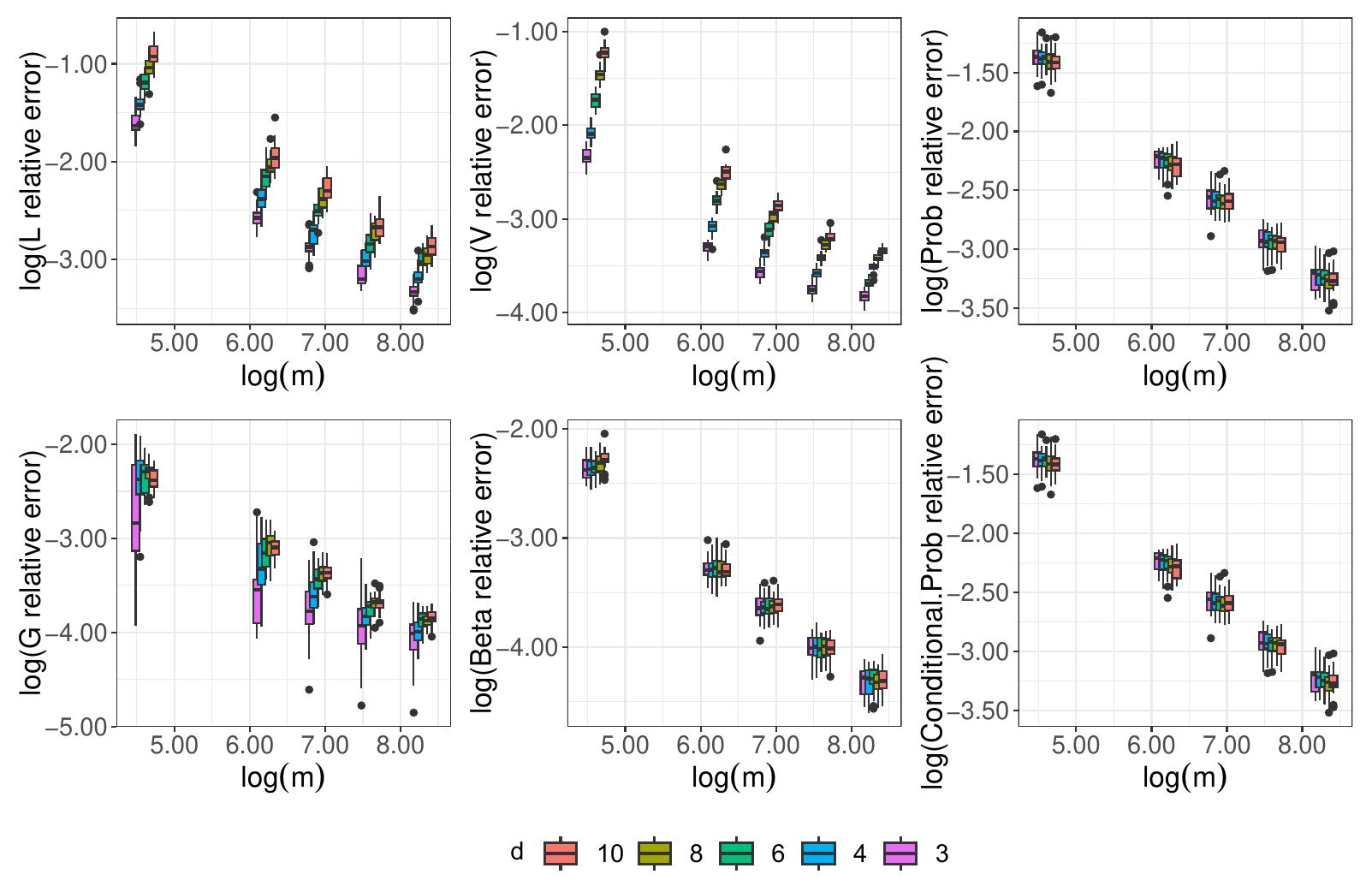}
	\caption{Relative errors under the NDPP model. Latent positions for nodes are uniformly generated with a variety dimensions $d$ and a variety number of hyperedges $m$ as $s = 3$ and  $n = 100$.}
	\label{Relative_error_change_d}
\end{figure}

To evaluate the estimation accuracy, we report relative estimation errors calculated as:
$$
\frac{\min _{O \in O_d, S \in H_{n}}\left\|\hat{V}-S V^* O\right\|_F}{\|V^{*}\|_F},\,\, \frac{\|\hat{G} - G^{*}\|_F}{\|G^{*}\|_F},\,\,
\frac{\|\hat{B} - B^{*}\|_2}{\|B^{*}\|_F},\,\,
\frac{\min _{S \in H_{n}}\left\|\hat{L}-S L^* S\right\|_F}{\|L^{*}\|_F}, $$
where $\hat{V}$, $\hat{G}$, $\hat{B}$ and $\hat{L}$ are estimated parameters. 
Also reported is the relative estimation error of marginal and pairwise conditional edge probabilities, calculated as 
$$
\operatorname{mean}\left(\left| \frac{\hat{P}_i - P^{*}_{i}} {  P^{*}_{i}}\right|\right), \,\, \operatorname{mean}\left(\left|\frac{\hat{P}_{j \mid i}-P^{*}_{j \mid i}}{P^{*}_{j \mid i}}\right|\right),
$$
where $P_i = P( i \in E)$ and $P_{j \mid i}=P(j \in E \mid i \in E)$. Figures \ref{Relative_error_change_d} and Figure S1 report the simulation results from 50 data replications under different distributions for $V_i$'s. 

%We first validate the theoretical results by showing the consistency of our algorithm when changing the number of hyperedges and the scaling level parameters. The hyperedges are sampled following the NDPP distribution. The sampling algorithm is adapted from the algorithm provided in \cite{han2022scalable}. The details are shown in Algorithm \ref{NDPP sampling}.  In Figure \ref{Relative_error_change_d}, we present the log relative errors with varying number of hyperedges from 100 to 4000 and  varying latent space vector dimensions $d$ from 3 to 10 when when the number of nodes is 100. It can be observed that the parameters and the probabilities are all consistent to the true values. The relative errors decrease as the number of hyperedges increases. Although larger number of dimensions $d$ will lead to slightly larger relative errors for parameter estimation on $V, G, B, L$, the relative errors for marginal probability and pairwise probability do not have significant differences, which demonstrates the effectiveness of our algorithm. 

\begin{figure}
\hspace{2em}
 \includegraphics[width=0.9\textwidth]{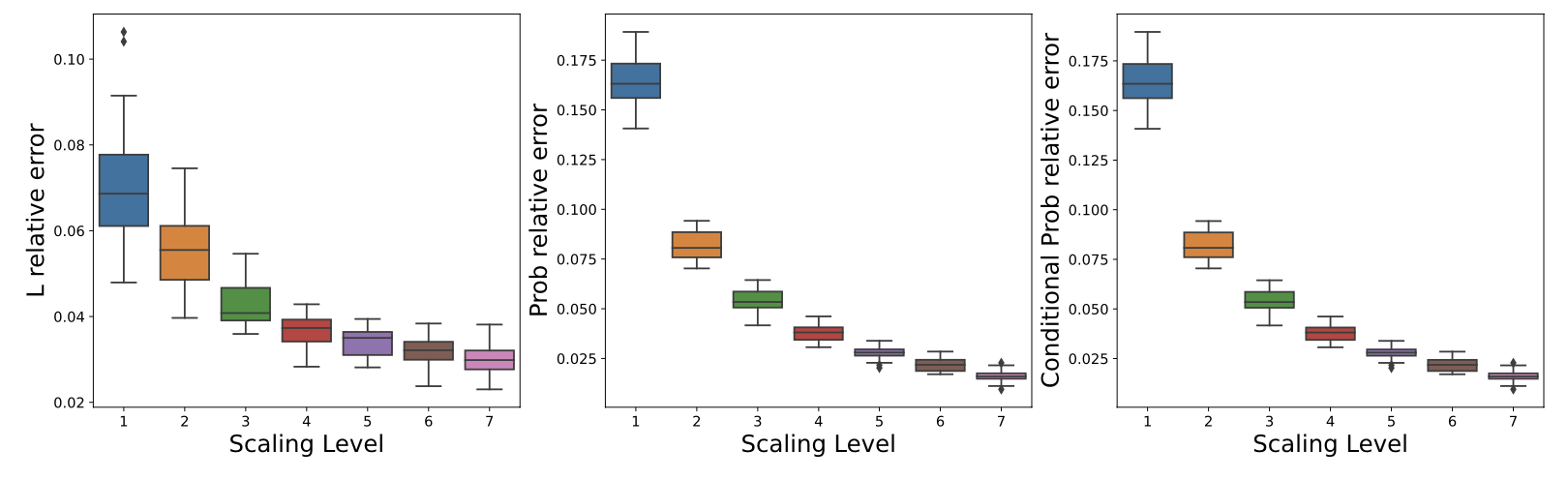}
	\caption{Relative errors under the NDPP model.  Latent positions for nodes are uniformly generated with a variety scaling levels $s$ as $d = 3, n = 100, m = 2000$. }
	\label{Relative_error_change_sparsity}
\end{figure}

Figure \ref{Relative_error_change_d} presents the log relative errors as the number of hyperhyedges $m$ varies from 1000 to 4000 and the latent vector dimension $d$ varies from 3 to 10. It is seen the estimated parameters and probabilities are close to the true values, and the estimation error decreases with $m$, consistent with our theoretical results. It can also be observed that the estimated marginal and conditional probabilities are not noticeably affected by $d$. Figure \ref{Relative_error_change_sparsity} shows how the relative errors vary with the scaling parameter $s$. It is seen that with a fixed $m$, the estimation error improves with the size of hyperedges.

Next, we compare the performance of DPP and NDPP models in two settings. In the first setting, the hyperedges are generated from the NDPP distribution, and in the second setting, the hyperedges are generated from the DPP distribution, using Algorithm 1 provided in \cite{kulesza2012determinantal}. In both settings, we estimate model parameter $V$, marginal probabilities and conditional probabilities using both the NDPP and DPP models. We set the number of nodes as $n=100$, the dimension as $d=3$, the scaling parameter as $s=3$. The number of hyperedges $m$ is set to vary from 100 to 4000.

\begin{figure}
\hspace{2em}
 \includegraphics[width=0.9\textwidth]{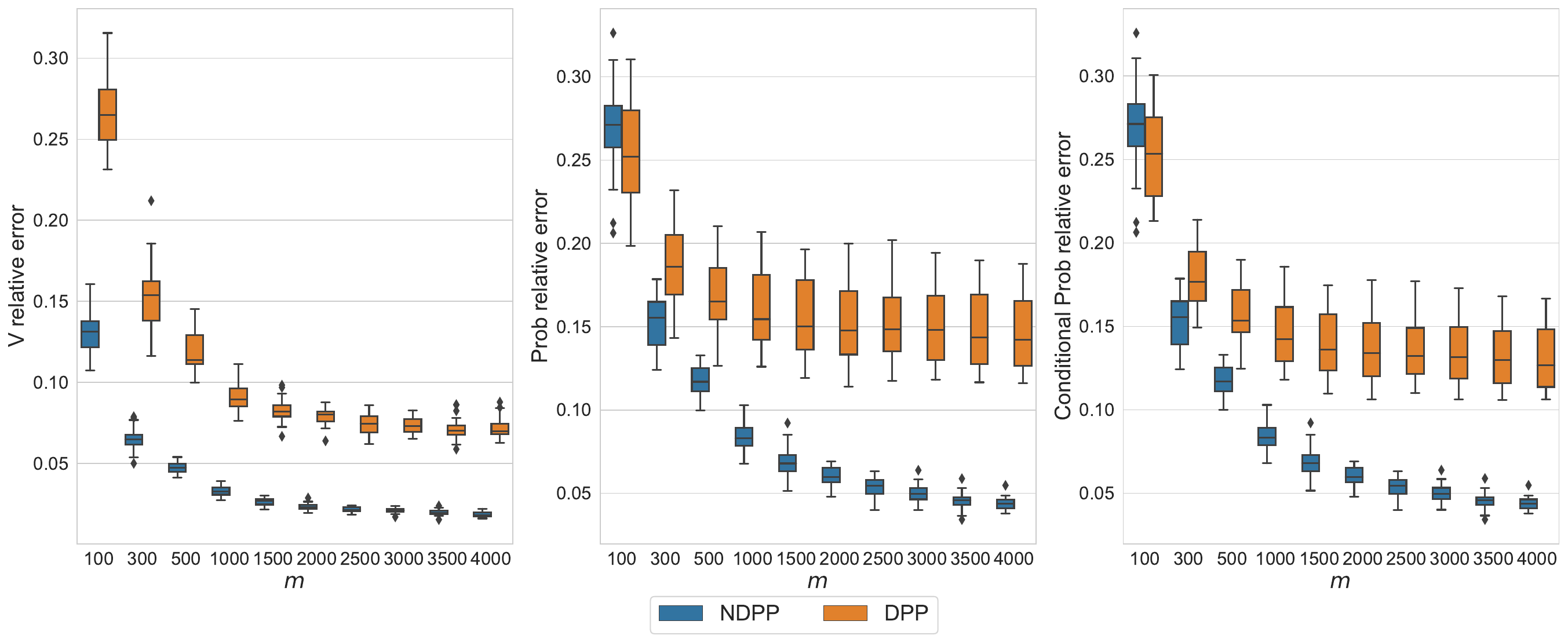}
	\caption{Comparison of relative errors under the NDPP and DPP model when the hyperedges are generated using NDPP. Latent positions for nodes are uniformly generated with a variety number of hyperegdes as $d = 3, s = 3, n = 100$. }
	\label{random_V_compare}
\end{figure}

Results from the first NDPP setting with uniformly distributed $v_i$'s are shown in Figures \ref{random_V_compare}. It is seen that relative errors from the NDPP model decrease with $m$ while the relative errors from the DPP model do not improve with $m$. Results from the second DPP setting with uniformly distributed for $v_i$'s are shown in Figures \ref{random_V_compare_DPP}. Additional results with vMF distributed $v_i$'s are shown in the supplement. It is seen that relative errors from the NDPP and DPP model are comparable, and both decrease with $m$. This demonstrates that NDPP is a more flexible model and can be used in more general situations. 

%The results are shown in Figure \ref{random_V_compare_DPP} and \ref{cluster_V_compare_DPP}. When we model the hypergraph using the NDPP model,  the relative errors still decrease to zero as the number of edges increases. We can further observe that the relative errors for both two models does not differ too much. It demonstrates that our NDPP model is more powerful and can be considered to use in more general situations. 

\begin{figure}
\hspace{2em}
 \includegraphics[width=0.9\textwidth]{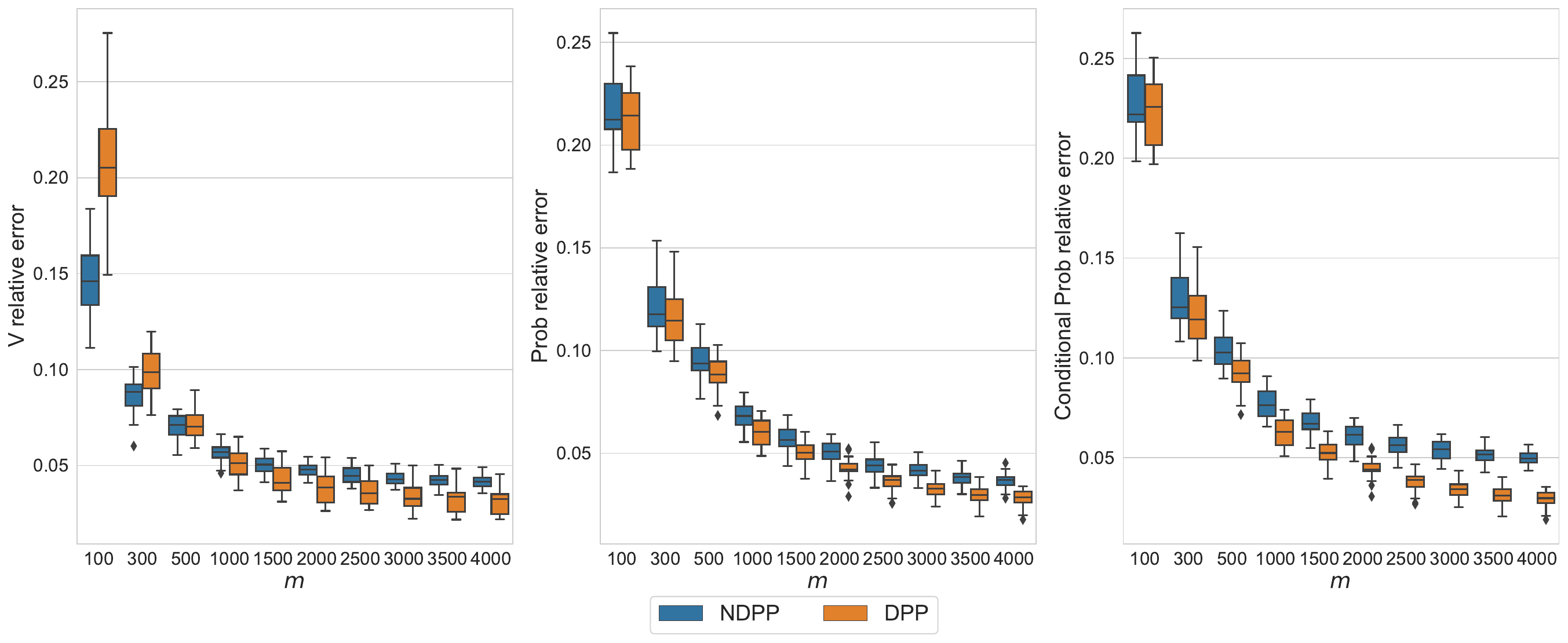}
	\caption{Comparison of relative errors under the NDPP and DPP model when the hyperedges are generated using DPP. Latent positions for nodes are uniformly generated with a variety  number of hyperegdes as $d = 3, s = 3, n = 100$. }
	\label{random_V_compare_DPP}
\end{figure}

\section{Real-world data examples}
We apply the proposed NDPP model to four distinct hypergraphs, focusing on the task of hyperedge prediction. The first hypergraph, named \textit{contact-high-school}, represents a human contact network constructed from interactions among high school students, recorded via wearable sensors. Here, the nodes are the students, and a hyperedge denotes the set of students recorded to be in close proximity to one another. The second hypergraph, named \textit{email-Eu}, is an email network constructed from a European research institution. In this network, the nodes are email addresses, and a hyperedge represents the set of accounts involved in an email, including both the sender and all recipients. The third hypergraph, named \textit{NDC-substances}, is a drug-related hypergraph constructed using data from the National Drug Code Directory. Nodes in this hypergraph correspond to individual substances (e.g., testosterone), and a hyperedge represents the collection of substances contained within a specific drug. The fourth hypergraph, named \textit{tags-math-sx}, is a tag network derived from the Mathematics Stack Exchange forum. Here, the nodes are tags (annotations), and a hyperedge denotes the set of tags assigned to a question on the forum. Further details on these hypergraphs can be found in \cite{benson2018simplicial}. Table \ref{tab: data information} provides a summary of these four hypergraphs.

On these four hypergraphs, we assess the performance of the DPP and NDPP models by evaluating their AUC (area under the ROC curve) and MPR (mean percentile rank) metrics. Hyperedges in each hypergraph are divided into training (80\%) and test (20\%) sets, and this procedure is repeated five times. The latent space dimension is fixed at 3 for the NDPP model. We also implemented NDPP and DPP by using the validation approach to select the latent space dimension. The results are consistent with our analyses, and are shown in Table S1 and Figure S3 in the supplement.

For AUC computation, we generate a set of random hyperedges matching sizes of true hyperedges in the test set, and calculate probabilities of both the randomly generated and true hyperedges using the fitted NDPP model. Labels are assigned as 1 for true hyperedges and 0 for the generated ones. The ROC curve then quantifies the model's ability to distinguish between true hyperedges and randomly generated ones. An AUC of 1 indicates perfect discrimination, whereas an AUC of 0.5 implies no better than random guessing. For MPR evaluation, one node from each test hyperedge is randomly masked as missing, and the NDPP model predicts the missing node's probability, conditional on the remaining nodes. The percentile rank of the true missing node is then calculated among all candidate nodes, and the MPR is derived by averaging these ranks across the test set. A percentile rank of 1 signifies the highest predictive accuracy, while a value of 0.5 indicates performance similar to random guessing.

\begin{table}[!t]
\centering

\begin{tabular}{ccccc}
\hline
\multirow{2}{*}{Data Set} & \multirow{2}{*}{Number of Nodes} & \multirow{2}{*}{Number of Hyperedges} & \multicolumn{2}{c}{Size of Hyperedges} \\
\cmidrule(lr){4-5}
& & & Mean $\pm$  Std & Range  \\
\midrule
contact-high-school & 242 & 1535 & 2.051 $\pm$ 0.241 & (2,4)\\

NDC-substances & 343 & 10068 & 3.795 $\pm$ 3.039 & (2,23)  \\
tags-math-sx & 302 & 15005 & 2.599 $\pm$ 0.802 & (2,5) \\
email-Eu & 153 & 25054 & 2.236 $\pm$ 1.139 & (2,17) \\\hline
\end{tabular}
\caption{Summary of the four hypergraphs. }
\label{tab: data information}
\end{table}

\begin{table}[!t]
\centering
\begin{tabular}{ccccc}
\hline
\multirow{2}{*}{Data Set} & \multicolumn{2}{c}{AUC} & \multicolumn{2}{c}{MPR} \\
\cmidrule(lr){2-3} \cmidrule(lr){4-5}
& NDPP & DPP & NDPP & DPP \\
\midrule
contact-high-school & \textbf{0.720} $\pm$ 0.015 & 0.680 $\pm$ 0.009 & \textbf{0.724} $\pm$ 0.009 & 0.690 $\pm$ 	0.007 \\
NDC-substances & \textbf{0.772} $\pm$ 0.005 & 0.747 $\pm$ 0.005 & \textbf{0.875} $\pm$ 0.005 & 0.832 $\pm$ 	0.004 \\
tags-math-sx  & \textbf{0.756} $\pm$	0.002 & 0.746 $\pm$	0.002  &  \textbf{0.812} 	$\pm$ 0.002 & 0.807 $\pm$	0.003 \\
email-Eu  & \textbf{0.682} $\pm$	0.002 & 0.634 $\pm$	0.002  &  \textbf{0.728}	$\pm$ 0.002 & 0.694 $\pm$	0.005 \\\hline

%\bottomrule
\end{tabular}
\caption{Performance comparisons for the NDPP and DPP model}
\label{tab: AUC and MPR}
\end{table}

\iffalse
\begin{figure}
\centering
% Row 2
\begin{subfigure}{.5\textwidth}
  \centering
  \includegraphics[width=.9\linewidth]{NDPP_contacth_AUC.pdf}
  \caption{AUC (contact-high-school Dataset) }

\end{subfigure}%
\begin{subfigure}{.5\textwidth}
  \centering
  \includegraphics[width=.9\linewidth]{NDPP_contacth_MPR.pdf}
  \caption{MPR (contact-high-school Dataset)}

\end{subfigure}
% Row 2
\begin{subfigure}{.5\textwidth}
  \centering
  \includegraphics[width=.9\linewidth]{NDPP_email_AUC.pdf}
  \caption{AUC (email-Eu Dataset) }

\end{subfigure}%
\begin{subfigure}{.5\textwidth}
  \centering
  \includegraphics[width=.9\linewidth]{NDPP_email_MPR.pdf}
  \caption{MPR (email-Eu Dataset)}

\end{subfigure}
% Row 3
\begin{subfigure}{.5\textwidth}
  \centering
  \includegraphics[width=.9\linewidth]{NDPP_NDC_AUC.pdf}
  \caption{AUC (NDC-substances Dataset)}

\end{subfigure}%
\begin{subfigure}{.5\textwidth}
  \centering
  \includegraphics[width=.9\linewidth]{NDPP_NDC_MPR.pdf}
  \caption{MPR (NDC-substances Dataset)}

\end{subfigure}
\caption{ Comparison of AUC and MPR for NDPP and DPP model}
\label{fig: AUC and MPR}
\end{figure}
\fi

\begin{figure}[!t]
\centering
% Row 1
\begin{subfigure}{.5\textwidth}
  \centering
  \includegraphics[width=.9\linewidth]{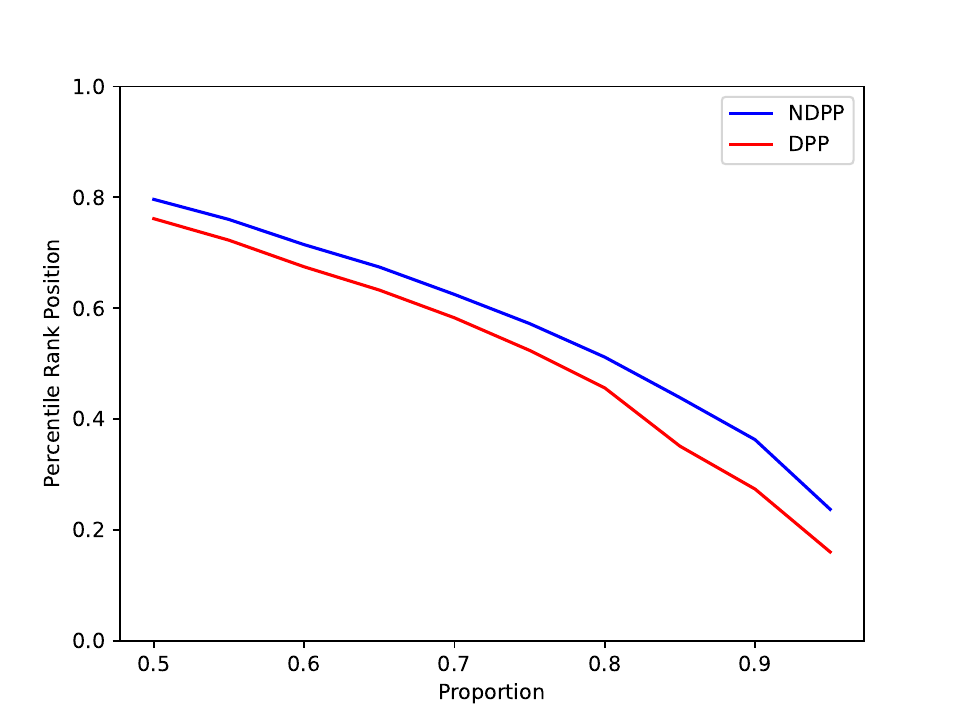}
  \caption{Email-Eu}

\end{subfigure}%
\begin{subfigure}{.5\textwidth}
  \centering
  \includegraphics[width=.9\linewidth]{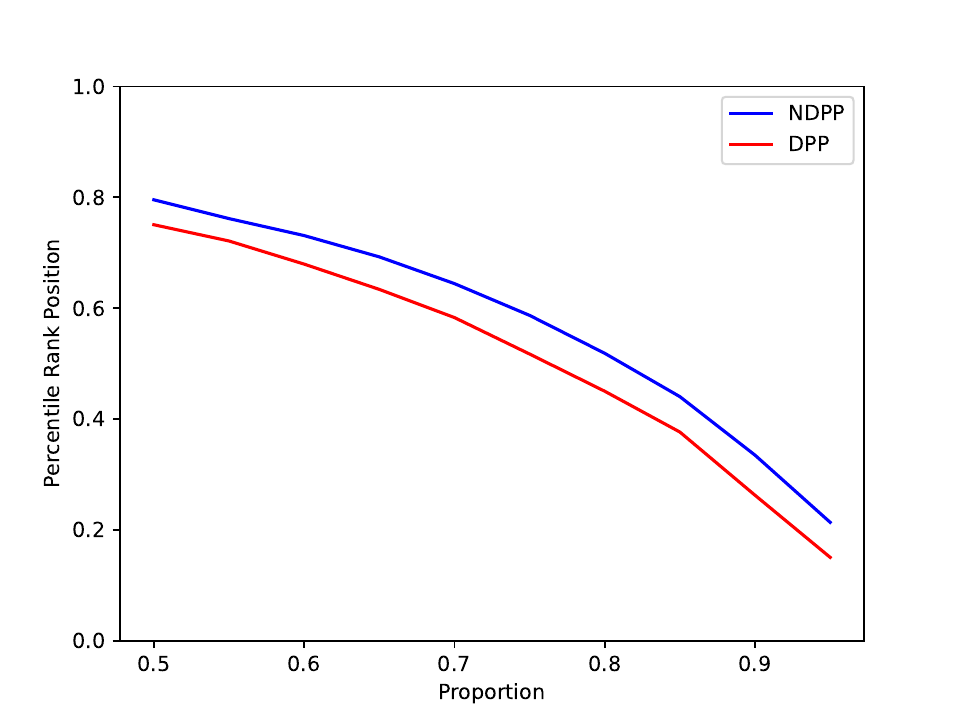}
  \caption{contact-high-school}

\end{subfigure}

% Row 2
\begin{subfigure}{.5\textwidth}
  \centering
  \includegraphics[width=.9\linewidth]{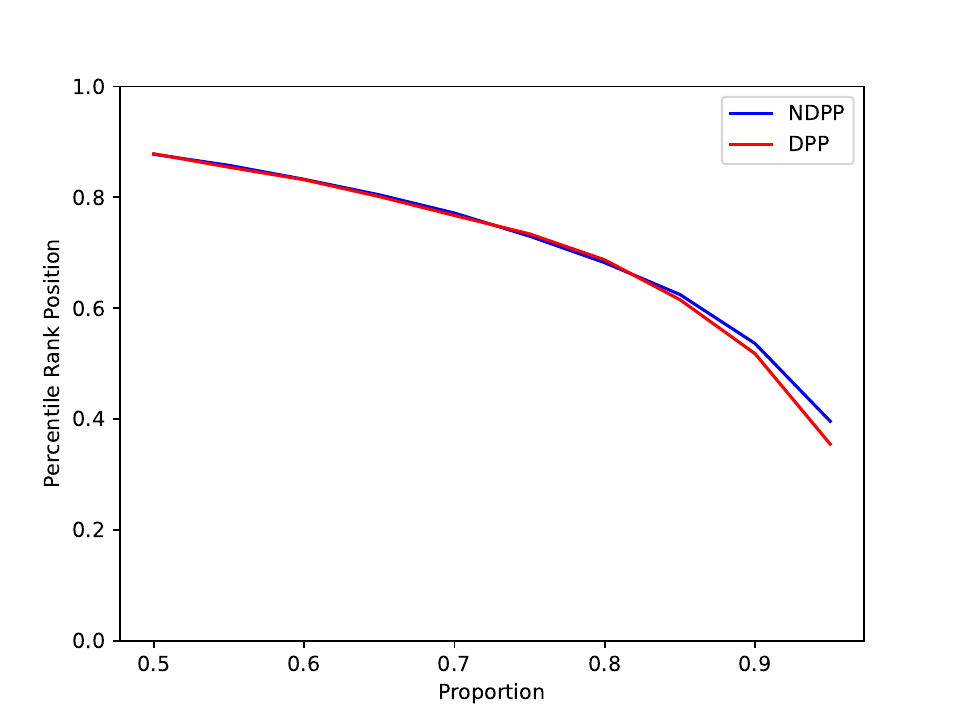}
  \caption{tags-math-sx}

\end{subfigure}%
\begin{subfigure}{.5\textwidth}
  \centering
  \includegraphics[width=.9\linewidth]{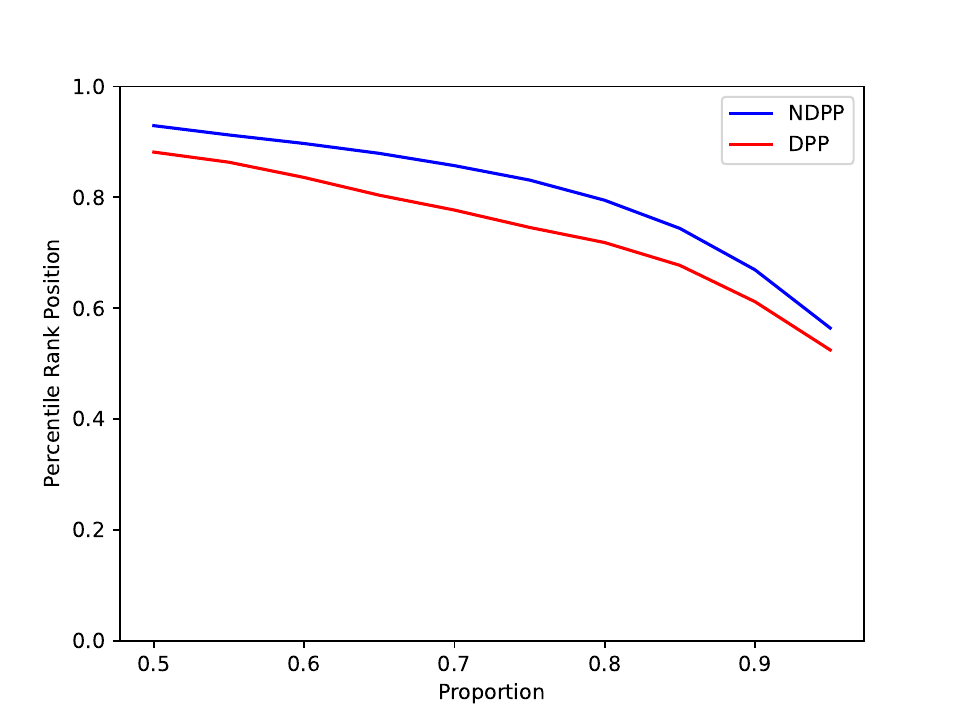}
  \caption{NDC-substances}

\end{subfigure}

\caption{Proportion of nodes with varying percentile ranks.}
\label{fig: PR curve}
\end{figure}

%\textcolor{blue}{temp notes:
%contact-high-school: NDPP d: 8,6,6,6,6 DPP: 4,3,4,4,3
%tag: NDPP d: 8,8,8,8,8 DPP: 4,4,8,8,4
%}

Table \ref{tab: AUC and MPR} shows the AUC and MPR results from the DPP and NDPP models, respectively. We also display the curve of percentile rank based on all nodes in Figure \ref{fig: PR curve}. For three out of four datasets, including contact-high-school, NDC-substances and email-Eu, the NDPP model outperforms the DPP model. This aligns with the inherent characteristics of the models. In contact-high-school, NDC-substances and email-Eu, similar nodes may appear more frequently in hyperdges, and the DPP model's node diversity assumption may not be appropriate. In contrary, in tags-math-sx, tags associated with a question tend to be more diverse. In this scenario, the DPP model performs similarly to the NDPP model.

%We also implement our model after choosing the latent space dimension for both NDPP and DPP model using a validation set. We split the dataset into training (60\%), validation (20\%) and test set (20\%). We choose $d$ from $[3,8]$ and select $d$ by the MPR performance on the validation set. The above process is repeated for 5 times. The results of AUC and MPR based on the 4 datasets is shown in Table \ref{tab: AUC and MPR (choose d)} and Figure \ref{fig: PR curve(choose d)} in the appendix. From the results, the NDPP model outperforms the DPP model significantly in terms of contact-high-school, NDC-substances and email-Eu datasets, which is consistent with our previous analyses. 

\section{Conclusion}    
In this paper, we propose a flexible model for non-uniform hypergraphs based on determinantal point processes, facilitating a significant reduction in the number of model parameters compared with tensor-based methods \citep{lyu2023latent}. 
Our model is the first non-uniform hypergraph model that can encourage node similarity within hyperedges. 
The consistency and asymptotic normality of the MLE estimates of the model parameters are also established. 

%However, the geometric explanation of our embedding is vague as it does not automatically reflect the distance between nodes by construction. Moreover, it remains open how we can interpret the estimated skew matrix. 
In the real-world data example part, we choose the dimension $d$ using the validation approach. It is worth to investigate how to select $d$ using a BIC-type criterion under our current framework. Additionally, extending the model to incorporate node-level covariates for more context-aware hyperedge formation is a worthwhile investigation. We leave these directions for future work.

\bibliographystyle{apalike}
\bibliography{draft.bbl}

\end{document}